\documentstyle[preprint,aps,epsf,floats]{revtex}    
\def\bq{\begin{equation}}
\def\eq{\end{equation}}
\def\ba{\begin{eqnarray}}
\def\ea{\end{eqnarray}}

\begin{document}
\thispagestyle{empty}
  
\renewcommand{\small}{\normalsize} 

\preprint{
\font\fortssbx=cmssbx10 scaled \magstep2
\hbox to \hsize{
\hskip.5in \raise.1in\hbox{\fortssbx University of Wisconsin - Madison} 
\hfill\vtop{\hbox{\bf CERN--TH/96--334}
            \hbox{\bf MADPH-96-978}} }
}
  
\title{\vspace{.5in}
Multijet Structure of High $E_T$ Hadronic Collisions
}

\author{D.~Rainwater$^1$, D.~Summers$^2$, and D.~Zeppenfeld$^1$}
\address{
$^1$Department of Physics, University of Wisconsin, Madison, WI 53706 \\[2mm]}
\address{
$^2$TH Division, CERN, CH-1211 Geneva 23, Switzerland 
}
\maketitle
\begin{abstract}
Multijet events at large transverse energy $(\sum E_T > 420$~GeV) and large 
multijet invariant mass ($m_{\rm jets}>600$~GeV) have been studied by the CDF
Collaboration at the Fermilab Tevatron. The observed jet multiplicity
distribution can be understood in a QCD inspired exponentiation model, 
in regions of phase space which require going beyond fixed order 
perturbation theory. 
\end{abstract}

%
%
\vspace*{1in}
\begin{flushleft}
\bf CERN--TH/96--334\\
December 1996
\end{flushleft}

\newpage
%
%

With the completion of run I at the Fermilab Tevatron, a sufficient amount 
of data has been collected to study rare hard scattering events, with 
jet transverse energies in the hundreds of GeV range. Such high $E_T$ 
jet events are interesting in their own right, as the discussion of a 
possible ``excess'' in the single jet $E_T$ spectrum 
above $E_T \approx 200$~GeV~\cite{CDFhighET} has shown. Another 
aspect is the fact that events with jets in this $E_T$ range, while rare
at the Tevatron, will be produced abundantly at the much higher energy of 
the CERN LHC, where they will constitute important backgrounds to searches
for new heavy degrees of freedom. Thus, the study of high $E_T$ QCD events
at the Tevatron allows one to develop strategies for new physics searches at
the LHC.

The aspect to be considered below is the multiplicity 
of additional soft jets which arise via the emission of soft gluons
in a hard scattering process. At the LHC, one application is the study
of weak boson scattering. In events like $qQ\to qQWW$, which are mediated 
by $t$-channel exchange of electroweak bosons, medium $E_T$ jets 
in the central region are a rare occurrence. This is a consequence of
color coherence between initial and final state radiation which leads
to gluon emission mainly between the forward scattered quarks and the 
beam directions~\cite{dkt}. At the same time the modest transverse 
momentum of the scattered quarks severely limits the transverse momentum 
of emitted gluons~\cite{DZ}. Typical background 
events like $t\bar t\to bW^+\bar{b}W^-$ or $q\bar{q}\to W^+W^-$, on the
other hand,  show a large probability for QCD radiation in the central 
region, with gluon transverse momenta sufficient to produce visible 
jets~\cite{bpz}. As a result, a central jet veto is quite effective 
for background suppression~\cite{bchz,CMS-ATLAS}. 

In applications like these it is necessary to correctly model the
angular distribution of emitted minijets and to reliably determine 
the hard scales which govern their transverse momenta. These features 
are automatically included by using full tree level
QCD matrix elements.  For an effective jet veto, however, one is interested
in the phase space region where the probability of extra QCD radiation
becomes of order unity, and this is exactly the region where fixed order 
perturbation theory ceases to be applicable. In Refs.~\cite{bpz,rsz}
it was suggested to use a soft gluon exponentiation model to extend the
perturbative calculation into this region of large minijet multiplicities.
In the following we show that existing data on multijet events at the 
Tevatron~\cite{CDFmultijet,D0multijet} allow one to test and refine this 
model for multiple minijet emission. In particular one can experimentally
determine ``good'' choices for the factorization and renormalization scales,
information which can then be used to more reliably predict minijet emission
in other processes, at the Tevatron or at the LHC. 

Consider dijet production in $p\bar p$ collisions at the Tevatron. The
next-to-leading order (NLO) QCD corrections to the cross section for 
producing two or more jets, $\sigma_{2,{\rm incl}}$\footnote{
Here and in the following, the ``cross section for n-jet inclusive 
events'', $\sigma_{n,\rm incl}=\sum_{k\geq n} \sigma_{k\;\rm jets}$,
directly corresponds to the rate of such events. No jet 
multiplicity factor is included in its definition.
}, is available in the form of a full NLO Monte Carlo program, 
JETRAD~\cite{ggk}.
The same program also provides the tree level cross section, $\sigma_3$, 
for three-jet production, within an arbitrary phase space region. In the 
following we use the CDF cone algorithm to define jets~\cite{CDFconealg}, 
with a radius of $R_0=0.7$ in pseudorapidity--azimuthal angle space; 
a cluster of partons of transverse energy $E_T > 20$~GeV 
and pseudorapidity $|\eta|< 4.2$ is defined as a jet. (Variations of 
the $E_T$ threshold will be considered later.) Following the 
analysis of the CDF Collaboration~\cite{CDFmultijet} 
we study two- and three-jet events with total jet transverse energy,
\bq
\label{eq:ETcut}
\sum E_T > 420~{\rm GeV}\; ,
\eq
and invariant mass of the multijet system, 
\bq
\label{eq:masscut}
m_{\rm jets} > 600~{\rm GeV}\; .
\eq
In addition the scattering angle~\cite{CDFkindef}, $\theta^*$, of the 
highest $E_T$ jet in the multijet center of mass frame must satisfy
\bq
\label{eq:thetacut}
|\cos\theta^*|< {2\over 3}\; .
\eq
Within these cuts $\sigma_{2,{\rm incl}}$ depends weakly on the 
minimal jet transverse energy since the two-jet inclusive events are 
largely defined by the two hardest jets, which  must have 
transverse energies $E_{T1},\; E_{T2}\approx \sum E_T/2 > 210$~GeV.
The three-jet cross section,
\bq
\sigma_3(E_{T,\rm min}) = 
\int_{E_{T,{\rm min}}}^\infty dE_{T3}{d\sigma_3\over dE_{T3}}\; ,
\eq
on the other hand, is a steeply falling function of the transverse
energy threshold, $E_{T,\rm min}$. 
For sufficiently low threshold one will eventually reach a region
with $\sigma_3(E_{T,{\rm min}})>\sigma_{2,{\rm incl}}$; clearly, the
interpretation of $\sigma_3$ as the cross section for either three jet
inclusive or three-jet exclusive events is not tenable in this region.
The unphysical relation $\sigma_3>\sigma_{2,\rm incl}$ is a sign that 
fixed order perturbation theory is breaking down and that multiple 
gluon emission needs to be resummed.

For small $E_{T,{\rm min}}$, soft gluon emission from the hard 
dijet production process will dominate, and, analogous to soft photon 
emission, one may show that this soft-gluon 
radiation approximately exponentiates when the soft gluons go 
unobserved\cite{expon}. Here we consider a phenomenological model 
which assumes that the analogy to multiple soft photon emission can be 
taken further, namely, that the probability $P_n$ for  
observing $n$ soft jets beside the two hard jets of the basic hard 
scattering event is given by a Poisson distribution,
\bq\label{eq:expon}
P_n(\bar n) = {\bar n^n\over n!}\; e^{-\bar n} \;,
\eq
with
\bq
\label{eq:nbar}
\bar n = \bar n(E_{T,\rm min}) = {1 \over \sigma_{2,\rm incl}}\; 
\int_{E_{T,\rm min}}^{\infty} dE_{T3}\; 
{d\sigma_3 \over dE_{T3}}\; .
\eq
We will call this model the ``exponentiation model'' in the 
following~\cite{bpz,rsz}.
The exponentiation model has a number of appealing features: 
\begin{enumerate}
\item{}
By construction the cross section for two jet inclusive events 
is given by $\sigma_{2,\rm incl}$.
\item{}
$\sigma_{2,\rm incl} P_0$ gives the correct Sudakov suppressed 
rate for two jet exclusive events~\cite{sudakov};
this is not surprising since $P_0$ is the probability for not seeing
any additional jets, and here soft gluon exponentiation can be proved.
\item{}
For sufficiently large $E_{T,\rm min}$, when
$\sigma_3(E_{T,{\rm min}})\ll\sigma_{2,{\rm incl}}$, the three jet
rate, {\em i.e.} the cross section for events with one soft jet,
is given by
\bq
\sigma_{2,{\rm incl}}\;P_1\;\approx\; \sigma_{2,{\rm incl}}\;\bar n \;
=\; \sigma_3(E_{T,{\rm min}})\; ,
\eq
and thus reproduces the perturbative result at 
${\cal O}(\alpha_s^3)$. From a perturbative point 
of view $\sigma_{2,{\rm incl}}P_1 = 
\sigma_3(E_{T,\rm min})(1+{\cal O}(\alpha_s))$, thus, 
$\sigma_{2,{\rm incl}}P_1$ has the same level of accuracy
as the tree level calculation of $\sigma_3(E_{T,\rm min})$.
\item{}
The full angular and transverse momentum information contained 
in the tree level result is retained in the estimated multijet 
emission probabilities.
\item{}
The probabilities $P_n$ all remain finite at small $E_{T,{\rm min}}$, 
with $0<P_n<1$. This renders the exponentiation model superior to
the use of {\em e.g.} $\sigma_3/(\sigma_{2,{\rm incl}}-\sigma_3)$ or
$\sigma_3/\sigma_{2,{\rm incl}}$ as an estimate for the ratio
of three-jet to two-jet exclusive or inclusive rates.
\item{}
Only the hard scattering cross section (here $\sigma_{2,{\rm incl}}$) 
and the cross section for emission of one additional soft parton 
(here $\sigma_3$) need to be known perturbatively. Thus, the model 
can easily be applied to more complicated processes~\cite{bpz,rsz}.
\end{enumerate}

\begin{figure}[t]
\vspace*{0.5in}            
\begin{picture}(0,0)(0,0)
\includegraphics{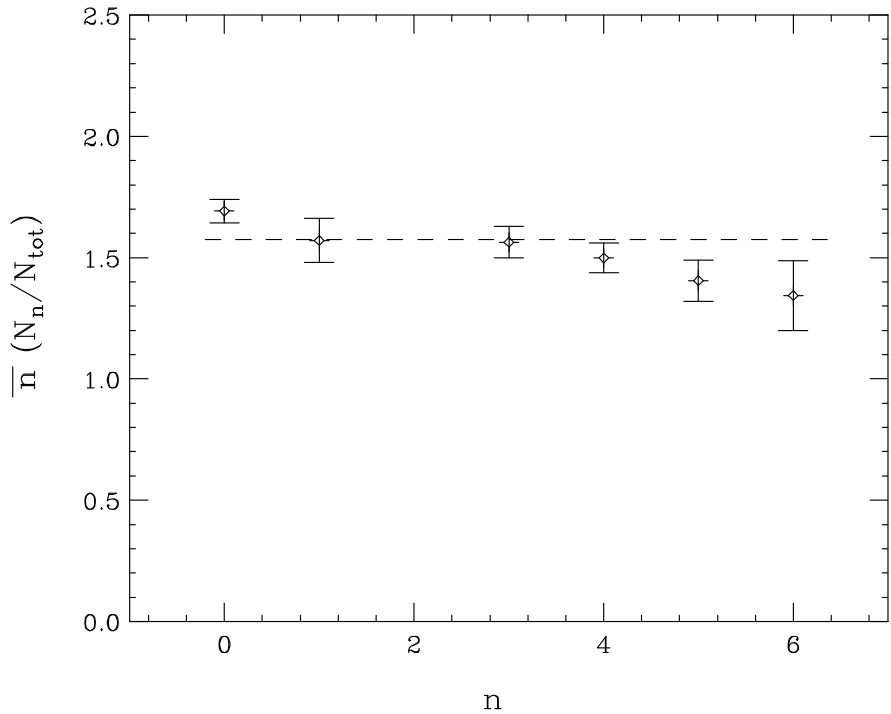}
\end{picture}
\vspace{9.0cm}
\caption{Poisson parameter $\bar n$ extracted from the fraction of
events with $n$ ``minijets'' in the CDF multijet 
sample~\protect\cite{CDFmultijet}, according to Eq.~(\protect\ref{eq:fn}). 
Errors are statistical only. A Poisson 
distribution of the minijet multiplicity would yield the dashed line 
at $\bar n=1.57$, the mean observed minijet multiplicity.
No solution to Eq.~(\protect\ref{eq:fn}) exists for $n=2$ since 
the observed fraction, $f_2=0.296\pm 0.011$, is larger than the maximal 
Poisson probability for two events, $P_2(\bar n=2)=0.271$. The dashed line
gives $P_2(\bar n=1.57)=0.257$ which is $3.5\sigma$ 
below the measured $f_2$.    
\label{fig:data}
}
\end{figure}

For the exponentiation model to be useful, it is necessary that minijet
multiplicities in hard scattering events at least approximately follow
a Poisson distribution. Recently the CDF collaboration has published 
results on multijet production within the cuts of 
Eqs.~(\ref{eq:ETcut}--\ref{eq:thetacut}). 
Out of a total of $N_{\rm tot}=1874$ events,
$N_0=345$ have exactly 2 jets with $E_T>20$~GeV, $N_1=612$ have 3 such jets, 
and the number of events with $n\geq 2$ minijets, in addition to the hard 
dijet system, are $N_2=554$, $N_3=250$, $N_4=88$, $N_5=21$, and 
$N_6=4$~\cite{CDFmultijet}. For a Poisson distribution the average 
multiplicity of minijets in the CDF event sample, $\left< n_{jets}-2\right> 
= 1.57\pm 0.03$ (where the error is statistical only), and the values
for $\bar n(N_n/N_{\rm tot})$ extracted from the fraction of events with 
a fixed number of minijets, via the relation
\bq
\label{eq:fn}
f_n = {N_n\over N_{\rm tot}} = P_n(\bar n)\; ,
\eq
should all agree. The results of these different extractions of $\bar n$
are compared in Fig.~\ref{fig:data}. A Poisson ansatz for the jet multiplicity
distribution works surprisingly well. For larger minijet multiplicities
($n\geq 4$) the observed rates fall more and more below a Poisson
distribution. This can be understood qualitatively, however, as an effect
of the reduced phase space (in $\eta$, $\phi$) which is available for 
additional minijets due to the finite cone size of jets. Also, one should note
that a systematic error of 15\% only on all $N_n$, 
(in addition to the statistical errors shown in  Fig.~\ref{fig:data})
would suffice to render a Poisson fit perfectly acceptable. 

Since the CDF data are well described by a Poisson distribution of minijet
multiplicities we may now compare the measured average multiplicity,
$\bar n = 1.57$ for $E_{T,\rm min}=20$~GeV, with the perturbative QCD 
prediction given in Eq.~(\ref{eq:nbar}). For this purpose we have calculated 
the total multijet rate, within the cuts of 
Eqs.~(\ref{eq:ETcut}--\ref{eq:thetacut}) and 10\%
Gaussian smearing of jet momenta~\cite{CDFmultijet} to simulate detector 
effects.  Using the JETRAD program~\cite{ggk}
we find $\sigma_{2,\rm incl}=33$~pb. Here we have used CTEQ 4HJ
parton distribution functions~\cite{cteq} and the factorization and the 
renormalization scale have been set to $\mu_F=\mu_R=\sum E_T/4$,
with $\alpha_s$ evaluated at two-loop order and $\alpha_s(M_Z)=0.116$
as required by the evolution of the parton distribution functions.

Since the 3-jet cross section $\sigma_3(E_{T,{\rm min}})$ is evaluated at 
tree level only, a variation of scales leads to substantial uncertainties
here. We have analyzed four different choices. (i) $\mu_R=\mu_F=\sum E_T/4$,
which is the same choice as for the 2-jet inclusive rate. Such a large 
factorization scale is not entirely physical, however. Collinear initial 
state radiation with transverse energies between $E_{T,\rm min}$ and the 
hard scale, $\sum E_T/4$, is generated explicitly in terms of the third jet.
When choosing a large factorization scale, this emission is considered twice,
in terms of the third jet and via the evolution of the parton distribution
functions. A factorization scale which is tied to the transverse energy 
of the third parton avoids such double counting. Similarly, a small
renormalization scale may better match the small parton virtualities 
which appear in the emission of soft gluons. This motivates 
the following set of scales: (ii) $\mu_R=\mu_F= E_{T,3}$,
where $E_{T,3}$ is the transverse energy of the softest of the three partons,
and, finally, $\mu_F=\xi E_{T,3}$ with an $\alpha_s^3$ factor in 
the calculation of $\sigma_3$ which is given by
\bq 
\alpha_s^3 = \prod_{i=1}^3 \alpha_s(\xi E_{T,i})\; ,
\eq
{\em i.e.} the transverse energy of each final state parton is taken as 
the relevant scale for its production. Here we have used overall scale
factors (iii) $\xi=1$ and (iv) $\xi=1/2$.

\begin{figure}[t]
\vspace*{0.5in}            
\begin{picture}(0,0)(0,0)
\includegraphics{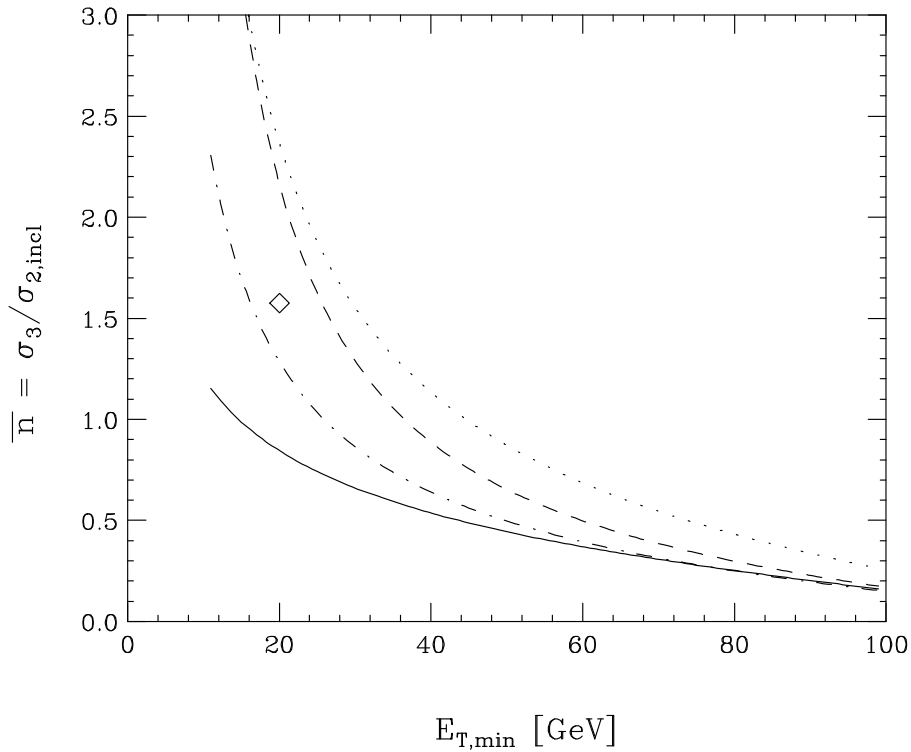}
\end{picture}
\vspace{9.0cm}
\caption{Ratio of the tree level 3-jet cross section to the NLO
cross section for 2-jet inclusive events within the CDF acceptance
cuts~\protect\cite{CDFmultijet}. The cross section 
ratio $\bar{n} = \sigma_3(E_{T,{\rm min}})/\sigma_{2,{\rm incl}}$,
with $\sigma_{2,{\rm incl}}=33$~pb, is shown as a function of the 
transverse energy threshold, $E_{T,{\rm min}}$, of the third jet.
Results are given for four different 
scale choices in $\sigma_3$: $\mu_R=\mu_F=\sum E_T/4$ (solid line),
$\mu_R=\mu_F= E_{T,3}$ (dashed line), and $\mu_F=\xi E_{T,3}$, 
$\alpha_s^3 = \prod_{i=1}^3 \alpha_s(\xi E_{T,i})$ with a scale factor
$\xi=1$ (dash-dotted line) and $\xi=1/2$ (dotted line). The CDF value for
the average minijet multiplicity, $\bar{n}=1.57$, is given by the diamond.
\label{fig:QCD}
}
\end{figure}

Results for these four scale choices are shown in Fig.~\ref{fig:QCD}.
For $E_{T,\rm min}=20$~GeV, estimates for the mean minijet multiplicity 
$\bar n(E_{T,\rm min}) = \sigma_3(E_{T,\rm min})/\sigma_{2,\rm incl}$ 
vary between 0.8 and 2.2 and thus bracket the CDF value of 1.57. From
Fig.~\ref{fig:QCD} it is clear, however, that a scale choice tied
to the transverse energy of the soft jets is preferred by the data.    
The single CDF data point is not sufficient, of course, to optimize 
the scale choice: an analysis of the observed minijet multiplicity 
as a function of the minimal jet transverse energy would be needed.

We conclude that the exponentiation model provides a good 
description of existing 
data on minijet multiplicities in hard scattering events at the Tevatron.
For a further quantitative comparison, the $E_{T,\rm min}$ dependence of 
the multijet rates would be most useful. The freedom in choosing the 
renormalization and factorization scales for the tree level 3-jet
cross section should allow for a reasonable parameterization of the 
data, as a function of $E_{T,\rm min}$ and the parameters of the hard 
event, provided the approximate Poisson multiplicity distribution of 
jets remains valid beyond the phase space region probed in 
Ref.~\cite{CDFmultijet}. Finally we note that a comparison with tree level
results for 4 or 5 jet cross sections~\cite{NJETS}, while complementary,
will be subject to very large theoretical uncertainties. Just like in  the
exponentiation model, there is a strong scale dependence of the tree 
level $n$-jet cross sections $(\sigma_{n\;\rm jets}\sim \alpha_s^n(\mu_R))$.
In addition one needs to use the tree level cross sections at the edge of 
the validity range of fixed order perturbative QCD and similar to the case 
of the 3-jet cross section discussed before, calculated $n$-jet rates 
will become unphysically large for sufficiently low $E_{T,\rm min}$. The 
exponentiation model addresses this last problem and should, thus, 
provide a valuable alternative in analyzing multijet emission in hard 
scattering events.

\acknowledgements
Useful discussions with M.~Seymour are gratefully acknowledged.
This research was supported in part by the 
University of Wisconsin Research Committee with funds granted by the 
Wisconsin Alumni Research Foundation and in part by the U.~S.~Department 
of Energy under Contract No.~DE-FG02-95ER40896.

\newpage             

\end{document}